\documentclass[pre, twocolumn, showpacs, amsmath, amssymb]{revtex4}

\usepackage[dvips]{graphicx}
\usepackage{amssymb} 

\newcommand{\cf}{{\it cf.} }
\newcommand{\diff}{{\mathrm d}}
\newcommand{\dGB}{\overline{\delta G}}
\newcommand{\dGM}{\delta G_\mathrm{M}}
\newcommand{\Ea}{E_{\mathrm a}}

\newcommand{\eg}{{\it e.g.}, }
\newcommand{\eq}[1]{eq.~(\ref{#1})}

\newcommand{\fig}[1]{figure~\ref{#1}}
\newcommand{\Fig}[1]{Figure~\ref{#1}}

\newcommand{\gs}{\bar g _{\sign (\bar g)}}
\newcommand{\ie}{{\it i.e.}, }
\newcommand{\pimed}{\pi_\mathrm{m}}
\newcommand{\piI}{\pi_{\mathrm I}}
\newcommand{\piA}{\pi_{\mathrm A}}

\newcommand{\PG}{P_\mathrm{G}}
\newcommand{\oper}[2]{\mathrel{\mathop{\kern 0pt#1}\limits_{#2}}}
\newcommand{\sign}{\mathrm{sign}}

\begin{document}

\title{Large phenotype jumps in biomolecular evolution}

\author{F. Bardou}%
\affiliation{IPCMS, CNRS and Universit\'e Louis Pasteur, 23 rue
du Loess, BP 43, F-67034 Strasbourg Cedex 2, France.}

\author{L. Jaeger}%
\affiliation{Chemistry and Biochemistry Department,
University of California, Santa Barbara,
Santa Barbara, CA 93106-9510}

\begin{abstract}
By defining the phenotype of a biopolymer by its active
three-dimensional shape, and its genotype by its primary sequence, we
propose a model that predicts and characterizes the statistical
distribution of a population of biopolymers with a specific phenotype,
that originated from a given genotypic sequence by a single mutational
event. Depending on the ratio $g_0$
that characterizes the spread of potential energies of
the mutated population with respect to temperature, three different 
statistical regimes have been identified.
We suggest that biopolymers found in nature are in a critical regime with
$g_0 \simeq 1-6$, corresponding to a broad, but not too broad,
phenotypic distribution resembling a truncated L\'evy flight. Thus the
biopolymer phenotype can be considerably modified in just a few
mutations. The proposed model is in good agreement with the experimental
distribution of activities determined for a population of single 
mutants of a group I ribozyme.
\end{abstract}


\maketitle

\section{Introduction}
The biological function (or phenotype) of a biopolymer, 
such as a ribonucleic acid (RNA) 
or a protein, is mostly determined by the three-dimensional 
structure resulting from the folding of linear sequence of 
nucleotides (RNA) or aminoacids (proteins) that specifies a genotype.
Generally, a natural biopolymer sequence (or genotype) 
codes for a specific two-dimensional or three-dimensional structure
that defines the biopolymer activity. But 
one sequence can simultaneously fold in several metastable 
structures that can lead to different phenotypes. 
Thus, random mutations of a sequence induce random changes of the 
metastable structure populations, which generates a random walk of 
the biopolymer function. Understanding this phenotype random walk 
is a basic goal for "quantitative" biomolecular evolution.
	
The statistical properties of RNA secondary structures considered 
as a model for genotypes have been investigated in depth in the recent 
years \cite{Fon2002}. The neutral network
concept \cite{Kim1968,Kim1983}, \ie the notion of a set of sequences, 
connected through point mutations, having roughly the same phenotype, has been 
shown to apply to RNA secondary structures. Thus, by drifting 
rapidly along the neutral network of its phenotype, a sequence may come 
close to another sequence with a qualitatively different phenotype, 
which facilitates the acquisition of new phenotypes through random 
evolution. Moreover, in the close vicinity of any 
sequence with a given structure, there exist sequences with nearly 
all other possible structures \cite{SFS1994},
as originally proposed in immunology \cite{PeO1979}.
Thus, even if the sequence space is much too vast to be explored through 
random mutations in a reasonable time (an RNA with 100 bases only has 
$10^{60}$ possible sequences), the phenotype space itself may be 
explored in a few mutations only, which is what matters biologically. 
These ideas have been brought into operation in a recent experiment
\cite{ScB2000} showing that a particular RNA sequence, catalyzing a 
given reaction, can be transformed 
into a sequence having a qualitatively different activity, using a
small number of mutations and without ever going through inactive steps. 

This paper investigates the phenotype space exploration at an
elementary level by studying the statistical distribution of a population
of biopolymers in a specific three-dimensional shape, that originated
from a given genotypic sequence by a single mutational event. 
It complements studies of 
the evolution from one structure to another structure \cite{FoS1998}, 
that consider only the most stable structure for each 
sequence and neglect the 
thermodynamical coexistence of different structures for the same sequence.
It also provides more grounds to the recent work
that suggests that RNA molecules with novel phenotypes evolved
from plastic populations, \ie populations folding in several structures,
of known RNA molecules \cite{AnF2000}.
It is experimentally evident, for instance in \cite{ScB2000}, that some 
mutations change the biopolymer chemical activity by a few percents 
while other mutations change it by orders of magnitude. 
This is not unexpected since, 
depending on their positions in the sequence, some residues have a dramatic
influence on the 3D conformation while others hardly matter.
Thus, the function random walk statistically resembles 
a L\'evy flight \cite{SZF1995,KPS1999,BBA2002} presenting jumps 
at very different scales. 
The respective parts of gradual changes and of sudden jumps in biological
evolution is a highly debated issue. While the gradualist point of view 
has historically dominated, evidences for the presence of jumps have accumulated
at various hierarchical levels from paleontology \cite{ElG1972}, 
to trophic systems, chemical reaction networks and
neutral networks and molecular structure \cite{FoS1998}.
The jump issue will be treated here by studying the statistical 
distribution describing the phenotype effects of random mutations of 
a biopolymer genotype. 

To address the question of the statistical effects of random
mutations of functionally active biopolymers,
we propose a model inspired from disordered systems physics that naturally
predicts the possibility of broad distributions of activities 
of randomly mutated biopolymers.
With two energy parameters describing the polymer energy landscape, this 
models is shown to exhibit a variety of behaviors and to fit experimental
data. Natural biopolymers are in a critical regime, related 
to the activity distribution broadness, in which a single mutation may have 
a large, but not too large, effect.

\section{Physical model of shape population distribution}

\begin{figure}
\includegraphics[scale=1.0,angle=0]{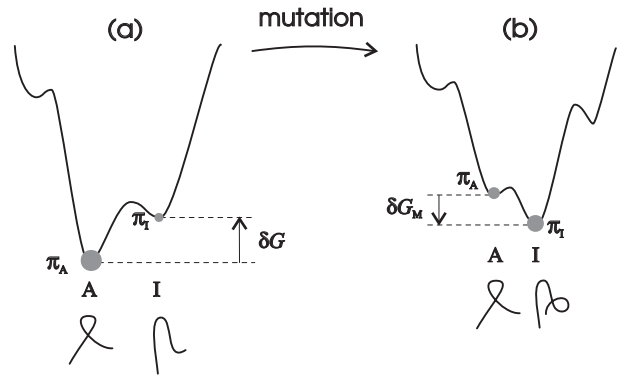}
\caption{Schematic representations of the molecular energy landscapes.
(a) For the non-mutated molecule. (b) For the mutated molecule. Only the 
two lowest energy conformations, A (active) and I (inactive), 
are taken into account. Their 3D conformations are indicated symbolically.
The shaded dots indicate the populations $\piA$ and $\piI$ at thermal 
equilibrium.}
\label{fig1}
\end{figure}

The most favorable conformational state of a biopolymer sequence with a given
biological activity is generally considered to be the most stable one
within the
sequence energy landscape. The ruggedness of the energy landscape might
vary depending on the number of other metastable, conformational states
accessible by the sequence. The typical energy spacing between these
states can be
small enough so that several states of low energy can be populated. For
simplicity, we will consider a sequence that is able to fold into its
two lowest energy conformational states, an active state A of 
specific biological function, and an inactive state I of 
unknown function \footnote{In our model, the inactive state may be 
replaced by an ensemble of inactive states with a given 
energy \cite{OLW1997}.}, but whose
energy is the closest to A's (higher or lower) (see \fig{fig1}). 
The differences between the free energies of the unfolded and folded 
states for A and I are denoted $\Delta G_\mathrm{A}$ and $\Delta G_\mathrm{I}$,
respectively.

A mutation, \ie a random change in the biopolymer sequence, modifies 
the biopolymer energy landscape so 
that $\Delta G_\mathrm{A}$ and $\Delta G_\mathrm{I}$ are transformed 
into $(\Delta G_\mathrm{A})_\mathrm{M}$ and
$(\Delta G_\mathrm{I})_\mathrm{M}$.
Note that the conformer state A of the mutant, its three dimensional shape, 
is the same as before whereas the conformer state I 
does not have to be the same as before. 
To take into account the randomness of the mutational process, 
the mutant free energy difference
$\delta G_\mathrm{M} \equiv (\Delta G_\mathrm{I})_\mathrm{M} - (\Delta
G_\mathrm{A})_\mathrm{M}$ is taken either 
with a Gaussian distribution:
\begin{equation}
P_\mathrm{G} (\dGM) \equiv \frac{1}{\sqrt{2 \pi} \delta G_0} \,
        e^{-(\dGM - \dGB)^2/ 2 \delta G_0^2} \, ,
\end{equation}
or with a two-sided exponential (Laplace) distribution:
\begin{equation}
P_\mathrm{e} (\dGM) \equiv \frac{1}{2 \, \delta G_0} \, e^{-|\dGM - \dGB| /
\delta G_0},
\end{equation}
where $\dGB$ is the mean of $\dGM$ and where $\delta G_0$ characterizes the
width of the distribution. These two energy distributions are commonly used 
for disordered systems \cite{DoH2002} and enable
us to cover a range of situations from narrow (Gaussian) to 
relatively broad (exponential) distributions.
Assuming thermodynamic rather than kinetic control, the populations 
$\piA$ and $\piI = 1-\piA$ of conformers A and I, respectively, are given by
Boltzmann statistics:
\begin{equation}
\piA = \frac{1}{1+e^{-\dGM / RT}},
\label{e3}
\end{equation}
where $R$ is the gas constant and $T$ is the temperature.

From the distributions of free energy differences and \eq{e3},
one infers the probability distributions 
$P_\mathrm{e \, or \, G} (\piA)$ of the population of
conformer state A after a mutation using $P_\mathrm{e \, or \, G} (\piA)=
P_\mathrm{e \, or \, G} (\delta G_\mathrm {M}) \times 
|\diff \delta G_\mathrm {M} / \diff \piA |$. 
For the Gaussian model, one obtains:
\begin{equation}
P_\mathrm{G} (\piA)  = \frac{\exp \left[ \frac{-\left( \ln \piA - \ln (1- \piA) - \bar g 
\right) ^2}{2 g_0^2} \right]}{\sqrt{2 \pi} g_0 \piA (1-\piA)} \, , 
\label{e5}
\end{equation}
where $\bar g \equiv \overline{\delta G}/(RT)$, $g_0 \equiv \delta G_0/ (RT)$. 
The ratio $g_0$ of the scale of energy fluctuations
and of the thermal energy appears frequently in the study of the anomalous
kinetics of disordered systems.
For the exponential model, one obtains:
\begin{subequations}
\label{e4}
\begin{equation}
P_\mathrm{e} (\piA)  = \frac{e^{-\bar g/g_0}}{2 g_0 \piA ^ 
{1-1/g_0} (1-\piA)^{1+1/g_0}} \; \mathrm{for} \; 
\piA \leq \pimed,
\end{equation}
\begin{equation}
P_\mathrm{e} (\piA)  = \frac{e^{+\bar g/g_0}}{2 g_0 \piA ^ 
{1+1/g_0} (1-\piA)^{1-1/g_0}} \; \mathrm{for} \;
\piA \geq \pimed,
\end{equation}
\end{subequations}
with the same definitions for $\bar g$ and $g_0$, and 
$\pimed \equiv \left( 1 + e^{-\bar g}\right) ^{-1}$ (median population of A).
Note that changing $\bar g$ into $-\bar g$ is equivalent to performing 
a symmetry on $P_\mathrm{e \, or \, G} (\piA)$ by replacing $\piA$ by $1-\piA$.

\section{Types of distributions}
\begin{figure}
\includegraphics[scale=0.33,angle=-90]{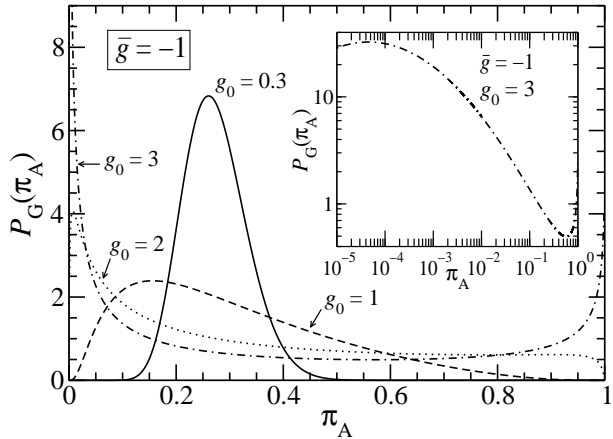}
\caption{Distributions $P_\mathrm{G} (\piA)$ of shape populations of mutated 
molecules for $\bar g = -1$. They are narrow and single peaked for small 
enough $g_0$ and broad and double peaked for large enough $g_0$. The transition 
from one to two peaks occurs at $g_0 \simeq 1.976$ in agreement with \eq{e6}.
Inset: logarithmic plot of $P_\mathrm{G} (\piA)$ for $g_0 = 3$ showing 
the broad character of the small $\piA$ peak.
}
\label{fig2}
\end{figure}

To analyze the different types of population distributions,
we focus for definiteness on the Gaussian model.
A qualitatively similar behavior is obtained for the exponential model.
\Fig{fig2} represents examples of $P_\mathrm{G} (\piA)$ for
the Gaussian model with $\bar g = -1$ and various
$g_0$'s.  The negative value of $\bar g$ implies that A is on average
less stable than I, and hence that $\piA$ is predominantly less than 50\%.
For small $g_0$, the distribution 
$P_\mathrm{G} (\piA)$ is narrow since the 
width $\delta G_0$ of the free energy distribution is small compared to $RT$ 
so that there are only small fluctuations of population around the most 
probable value. When the energy broadness $g_0$ increases,  
the single narrow peak first broadens till, when $g_0 \gtrsim 1.976$ 
it splits into two peaks,
close respectively to $\piA=0$ and to $\piA=1$. 
The broad character of $P_\mathrm{G} (\piA)$ can be intuitively understood
as a consequence of the non linear dependence of $\piA$ on $\dGM$.
Thus, when the fluctuations of $\dGM$ are larger than $RT$, \ie
when $g_0 \gtrsim 1$, the quasi exponential dependence
of $\piA$ on $\dGM$ (\eq{e3}) non linearly magnifies $\dGM$ fluctuations
to yield a broad $\piA$ distribution, 
even if $\dGM$ fluctuations are relatively small compared to the mean $\dGB$.
A similar mechanism is at work for tunneling in 
disordered systems \cite{DRB2002,RDB2003}.

\begin{figure}
\includegraphics[scale=0.33,angle=-90]{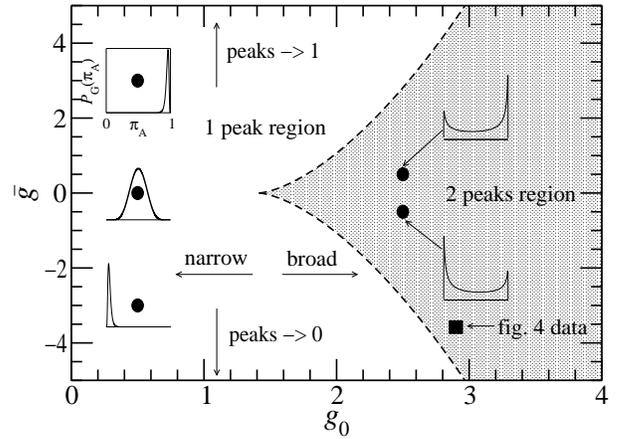}
\caption{Possible shapes of $P_\mathrm{G} (\piA)$. 
The shaded area indicates the two-peaks region. The dashed line gives the 
transition from one to two peaks (\cf \eq{e6}). Insets show examples of 
$P_\mathrm{G} (\piA)$ corresponding to the $g_0$ and $\bar g$ indicated by 
the black dots ($P_\mathrm{G}$'s not to scale).
The black square corresponds to the fit of \fig{fig4} data.}
\label{fig3}
\end{figure}

A global view of the possible shapes of $P_\mathrm{G} (\piA)$ is given in 
\fig{fig3}.
For any given $\bar g$, when increasing $g_0$ starting from 0, the
single narrow peak of $P_\mathrm{G} (\piA)$ first broadens 
then it splits into two peaks when $\bar g = \gs(g_0)$ with 
\begin{equation}
\bar g_\pm (g_0) \equiv \pm \left[ g_0 \sqrt{g_0^2 -2} + 
\ln \left( \frac{g_0- \sqrt{g_0^2 -2}}{g_0 + \sqrt{g_0^2 -2}}\right) \right].
\label{e6}
\end{equation}
(This expression results from a lengthy but straightforward
study of $P_\mathrm{G} (\piA)$.)
When $g_0$ increases further, these two peaks get closer to $\piA=0$ and 
to $\piA=1$ while acquiring significant tails (see section \ref{s7}).
For any given $g_0$, increasing $\bar g$ roughly amounts to
moving the populations $\piA$ towards larger values as expected since
larger $\bar g$'s correspond to stabler states A. However, distinct behaviours 
arise depending on $g_0$. If $g_0 < \sqrt{2}$, whatever the value of $\bar g$,
the distribution $P_\mathrm{G} (\piA)$ is always sufficiently narrow to  
present a single peak. If $g_0 > \sqrt{2}$, the distribution
$P_\mathrm{G} (\piA)$ is sufficiently broad to have two peaks when, 
furthermore, the distribution is not too asymmetric, which occurs for 
$\bar g \in [\bar g_-(g_0), \bar g_+(g_0)]$.
In short, depending on $\bar g = \delta \bar G/RT$, which characterizes 
mainly the peak(s) position, and on $g_0 = G_0/RT$, which characterizes 
mainly the distribution
broadness, the distributions $P_\mathrm{G} (\piA)$ are either unimodal
or bimodal, either broad or narrow. 
This variety of behaviors is reminiscent of beta distributions.

\section{From shape populations to catalytic activities}

Up to now, we have discussed the distribution $P(\piA)$ of the population
of a shape A that is functionally active. However, 
as far as it concerns biopolymers with
enzymatic functions, what is usually measured 
is a chemical activity $a$, \ie the product of a reaction rate $k$ for the
conformer A by the population $\piA$ of this conformer. The reaction rates 
are given by the Arrhenius law $k = k_0 e^{-\Ea/RT}$ where $k_0$ is a constant
and $\Ea$ is the activation energy. Thus, the chemical activity 
writes, using \eq{e3}:
\begin{equation}
a = k_0 e^{-\Ea/RT} \frac{1}{1+e^{-\dGM / RT}}.
\end{equation}
Random mutations may induce random modifications of $\Ea$, $\dGM$ or both.
Fluctuations of $\dGM$ have been treated above. One can 
introduce fluctuations of $\Ea$ in the same way. We do 
not do it here in details but present only the general trends.

The effects of adding an activation energy distribution in addition to the
free energy difference distribution are twofold. 
For small activities, the distribution $P(a)$ of chemical
activities is similar to the small $P(\piA)$ peak at small $\piA$. Indeed,
the reaction rate $k$ depends exponentially on $\Ea$, just as the population
$\piA$ depends exponentially on $\dGM$ when $\piA \ll 1$. Moreover, the product
of two broadly distributed random variables is also broadly 
distributed \footnote{With Gaussian distributions of $\dGM$ and $\Ea$, one 
can be more specific. Both $\piA$ and $k$ are then lognormally distributed
at small values. Thus, the product $a = k\piA$ is also lognormally distributed
\cite{RDB2003}.} with a shape similar to the one of $P(\piA)$. 
For large activities, on the other hand, $\piA$ and $k$ behave 
differently because $\piA$ is bounded by 1 while $k$ is unbounded. Thus, 
if the $k$ distribution is broad enough, the distribution of $a$ at large
$a$ may exhibit a broadened structure compared to the $\piA \simeq 1$
peak of $P(\piA)$. 

In summary, the distribution of chemical activities $P(a)$ is similar to the 
distribution of shape populations $P(\piA)$ when $P(\piA)$ presents a large
$\piA \simeq 0$ peak (conditions for this to occur are explicited in 
section \ref{s6}). Thus, by observing the shape of the $a \simeq 0$ peak
in the activity distribution $P(a)$,
one does not easily distinguish between activation energy dispersion,
which affects $k$, and free energy difference dispersion, which affects
$\piA$. On the other hand, at large $a$, $P(a)$ is differently 
influenced by activation 
energy dispersion and by free energy difference dispersion. 
The available experimental data (see section \ref{s5}) 
enables us to analyze precisely
$P(a)$ at small activities but not at large activities. Thus, for practical
purposes, it is not meaningful in this paper to consider a distribution
of activation energies on top of a distribution of free energy differences.
In the sequel, we will thus do as if only the distribution of free energies
was involved, stressing that similar effects can be obtained from a 
distribution of activation energies.

\section{Analysis of experimental data \label{s5}}
Comparison of the theoretical distributions of \eq{e4} and \eq{e5} with 
experimental data enables us to test the relevance of the proposed 
model. We have analyzed the measurements of the catalytic activities 
of a set of 157 mutants derived from a self-splicing group I 
ribozyme, a catalytic RNA molecule \cite{CEG1990}
(out of the 345 mutants
generated in \cite{CEG1990}, we only considered the 157 ones with
single point mutations).
The original "wild-type" molecule is formed of a conserved catalytic
core that catalyzes the cleavage of another part of the molecule
considered as the substrate.
The set of mutants is derived from the original ribozyme by performing 
systematically all single point mutations of the catalytic core, 
\ie of the part of the molecule that influences most the catalytic activity.
Nucleotides out of the core, that in general influence less the catalytic 
activity, are left unmutated. 
Thus, in our framework, this set of mutants can be seen as biased towards 
deleterious mutations. Indeed, mutations of the quasi optimized core 
are likely to
lead to much less active mutants, while mutations of remote parts are likely to
leave the activity essentially unchanged. 
If all parts of the molecule had been mutated, more neutral or quasi 
neutral mutations would have been obtained. Another point of view, 
which we adopt here is to consider the catalytic core as a molecule in itself,
on which all possible single point mutations have been performed.

The 157 measured activities are used to calculate a population distribution
with inhomogenous binning (\cf broad distribution). Two bins
required special treament: the smallest bin, centered in 0.5\%, 
contains 40 mutants with non measurably small activities 
($<1\%$ of the original activity); 
the largest bin,
centered in 95\%, contains the 6 mutants with activities larger than 90\%
of the original 'wild' RNA activity (the largest measured mutant activity is 
140\%). These two points, whose abscissae are arbitrary within an interval, 
are not essential for the obtained results. At last, 
as very few mutants have activities larger than the wild-type ribozyme,
the proportionality constant between activity and population is set by
matching a population $\piA=1$ to the activity of the wild-type ribozyme.

\begin{figure}
\includegraphics[scale=0.33,angle=-90]{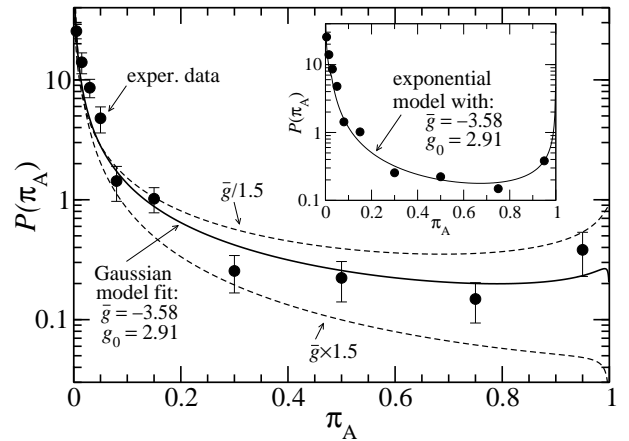}
\caption{Analysis of an experimental distribution of activities. Experimental
data are derived from \cite{CEG1990}. Error bars give the one standard 
deviation statistical uncertainty. The solid line is a two parameter fit
($\bar g$, $g_0$) to the model of Gaussian energy distribution. 
The dashed lines correspond to the same $g_0$ and modified $\bar g$'s, 
which enables to estimate the uncertainty on $\bar g$.
Inset: comparison of the data to the model of exponential energy distribution
($g_0$ and $\bar g$ are not fitted again but taken from the Gaussian model fit).
}
\label{fig4}
\end{figure}

The obtained distribution (see \fig{fig4}) has a large peak in $\piA \simeq 0$,
indicating that most mutations are deleterious, 
with a long tail at larger activities and a possible smaller peak in 
$\piA \simeq 1$. This non trivial shape is well fitted
by the Gaussian model of \eq{e5} 
with $\bar g = -3.6$ and $g_0 = 2.9$ (the uncertainty on these parameters
is about 50\%, see dashed lines in \fig{fig4}). One infers 
$\dGB \simeq -2.1$ kcal/mol and 
$\delta G_0 \simeq 1.7$ kcal/mol ($T=300$~K). 
The order of magnitude of these values is
compatible with thermodynamic measurements performed on
similar systems \cite{JWM1993,JMW1994,BrW1997,BMS1999}. 
This confirms the plausibility of the 
proposed approach. The inset of \fig{fig4} shows the population 
distribution in the {\it exponential} model with 
$\bar g$ and $g_0$ values taken from the {\it Gaussian} fit. The 
agreement with the experimental data is also quite good. Thus, the proposed
approach soundly does not strongly depend on the yet unknown shape details
of the energy distribution. 
Finally, one can estimate the broad character of the activity 
distribution from the statistical analysis of the experimental data.
Indeed, according, \eg to the Gaussian model fit, the typical, most probable,
population $\piA$ is found to be $\simeq 6\times 10^{-6}$ while 
the mean population is $\simeq 0.15$. Thus, the activity distribution 
spans more than four orders of magnitude.

\section{Coarse graining description: all or none features \label{s6}}
The variation of activity of a biopolymer upon mutation is
often described as an `all or none' process: mutations are considered 
either as neutral (the mutant retains fully its activity and 
$\piA \simeq 100$~\%) or as lethal (the mutant loses completely its activity 
and $\piA \simeq 0$~\%). Satisfactorily, a coarse graining description 
of the proposed statistical models exhibits such all or none regimes 
for appropriate $(\bar g, g_0)$ values, as well as other regimes.

\begin{figure}
\includegraphics[scale=0.33,angle=-90]{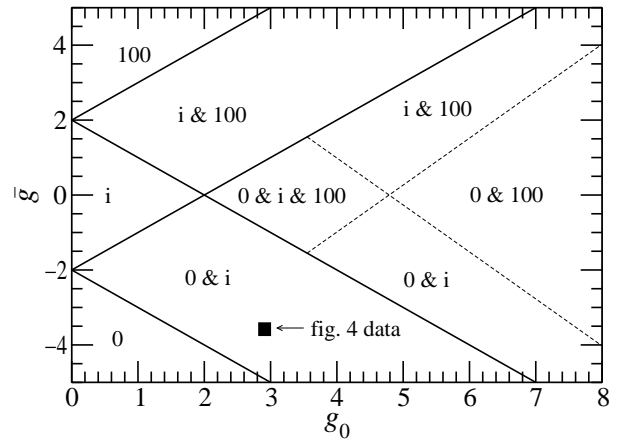}
\caption{Coarse graining features of the population distribution
$\PG(\piA)$ in the Gaussian model. In each region, the population ranges 
dominating the distribution have been indicated ($0$ for $\piA \leq 12\%$,
i for $12 \% \leq \piA \leq 88 \%$ and 100 for $\piA \geq 88 \%$).
}
\label{fig5}
\end{figure}

To obtain a quantitative coarse graining description,
we define the mutants with 'no' activity as those with population 
that has less than $12\ \%$ ($\simeq \piA ( \dGM = -2RT)$) in the 
A shape.
Their weight is
\begin{equation}
w_0 = \int_0 ^{12\%} \!\!\!\! P_\mathrm{e \, or \,  G} (\piA) \diff \piA =
\int_{-\infty}^{-2RT} \!\!\!\! P_\mathrm{e \, or \,  G} (\delta G) \ 
\diff \delta G .
\end{equation}
Similarly, the mutants with `full', respectively 'intermediate', activity
are defined as those with $\piA \geq 88 \%$, respectively 
$12 \% \leq \piA \leq 88 \%$, and their weight is 
$w_{100} = \int_{2RT}^{\infty} P_\mathrm{e \, or \,  G} (\delta G)
\ \diff \delta G$, respectively 
$w_{\mathrm i} = \int_{-2RT}^{+2RT} P_\mathrm{e \, or \,  G} (\delta G)
\ \diff \delta G$.
Taking for definiteness the Gaussian model leads to 
\begin{equation}
w_0 = \Phi \left( -\frac{2}{g_0} - \frac{\bar g}{g_0} \right)
\end{equation}
where $\Phi(u)= \int_{-\infty}^u e^{-t^2/2} \diff t / \sqrt{2\pi}$ 
is the distribution function of the normal distribution.
Similarly, one has $w_{\mathrm i} = \Phi[(2-\bar g)/g_0] - 
\Phi[-(2+\bar g)/g_0]$ and $w_{100} = 1 - \Phi[(2-\bar g)/g_0]$.
Approximate expressions for $\Phi(u)$ 
($\Phi(u) \simeq - e ^{-u^2/2}/(\sqrt{2 \pi} u)$ for $u\ll -1$, 
$\Phi(u) \simeq 1/2 + u /\sqrt{2 \pi}$ for $|u|\ll 1$ and 
$\Phi(u) \simeq 1 - e ^{-u^2/2}/(\sqrt{2 \pi} u)$ for $u\gg 1$) 
give the regimes in which each weight $w$ is
negligible ($w \ll 1$), dominant ($1-w \ll 1$) or in between.
For instance, $w_0$ is negligible for $\bar g > g_0 -2$, dominant for 
$\bar g < -g_0 -2$ and intermediate for $-g_0 -2 < \bar g < g_0 -2$.
These inequalities indicate the transition from one regime to another. 
To be strictly in one regime requires typically that $\bar g / g_0$ is 
larger or greater than 1 from the corresponding criterion, \eg
$w_0$ is strictly negligible when $\bar g / g_0 > 1 + (g_0 -2)/g_0$
The transitions from one regime to another one are in general 
exponentially fast (solid lines in \fig{fig5}). However, in the region 
($g_0>2$, $|\bar g| < g_0-2$), the transitions from one regime to another 
one are smooth (dashed lines in \fig{fig5}) since, in this region, 
the weights vary slowly, \eg $w_{\mathrm i} \simeq 4/(g_0\sqrt{2\pi})$.

The resulting coarse graining classification of $\PG(\piA)$ is represented 
in \fig{fig5}.
The `all or none' behaviour, denoted `0 \& 100', appears in the region
$g_0 \gtrsim 6/\sqrt{\pi/2}$  and $|\bar g| \lesssim \sqrt{\pi/2}g_0 -6$
as the result of a large dispersion of energy differences 
associated to a moderate average energy difference. We note that all possible
types of distributions are actually present in this model: probabilities
concentrated at small, intermediate or large values (0, i or 100); probabilities
spread over both small and intermediate (0 \& i), both small and large 
(0 \& 100, all or none) or both intermediate and large (i \& 100) values; 
probabilities spread over small, intermediate and large values at the same time (0 \& i \& 100).
The coarse graining classification of \fig{fig5} complements the number of 
peaks classification of \fig{fig3} without overlapping it. Indeed, there exist
parameters $g_0$ and $\bar g$ for which, \eg two peaks coexist but one of 
these peaks has a negligible weight. Thus the presence of a peak 
is not automatically
associated to a large weight in the region of this peak.

\section{Zooming in the $\piA \simeq 0$ peak: long tails \label{s7}}
To go beyond the coarse graining description, we zoom in 
the $\piA \simeq 0$ peak. As shown in the inset of \fig{fig3},
the small activities, 
labelled as `no activity' in a coarse graining description,
actually consist of non zero activities with values scanning 
several orders of magnitude. This can be analyzed quantitatively, \eg
in the Gaussian model. For $\piA \simeq 0$, the activity distribution
given by \eq{e5} is quasi lognormal:
\begin{equation}
\PG(\piA) \simeq \frac{1}{\sqrt{2\pi} g_0 \piA} \exp \left[ 
\frac{-(\ln \piA - \bar g)^2}{2 g_0^2}\right] \ .
\end{equation}
Thus, $\PG(\piA)$ has as a power law like behavior \cite{RDB2003,MoS1983}:
\begin{equation}
\PG(\piA) \simeq \frac{1}{\sqrt{2 \pi} g_0 \piA} \quad \mathrm{for} \quad
e^{\bar g - \sqrt{2} g_0} \lesssim \piA \lesssim e^{\bar g + \sqrt{2} g_0},
\label{e10}
\end{equation}
in the vicinity of the lognormal median $e^{\bar g}$.
This corresponds to an extremely long tailed distribution,
since $1/\piA$ is not even normalizable. 
It presents the peculiarity that, for $a$ and $a+1$
belonging to $[\bar g - \sqrt{2} g_0,\bar g + \sqrt{2} g_0]$,
the probability to obtain a population $\piA$ of a given order 
of magnitude $a$, \ie $\piA \in [e^a,e^{a+1}]$, does not depend 
on the considered ordered of magnitude $a$, since 
\begin{equation}
\int_{e^a}^{e^{a+1}} \PG(\piA) \diff \piA \simeq \mathrm{const}.
\end{equation}
Thus, if a living organism has to adapt the chemical activity of one of
its biopolymer constituents, it can explore several order of magnitude of
activity by only few mutations within the biopolymer.
The activity changes mimic a L\'evy flight \cite{BoG1990} as revealed, \eg 
by the experimental data in \cite{ScB2000}. The large activity changes 
will raise self-averaging issues \cite{RDB2003} that will add up to those
generated by correlations along evolutionary paths \cite{BPR2002}

Three broadness regimes corresponding to three evolutionary regimes can 
be distinguished. If $g_0$ is very large, the mutant activities span a very
large range. This regime might be globally lethal because, in most cases, the 
mutant activity will be either too low or too large to be biologically useful.
However, under conditions of intense stress, the large variability might allow 
the system to evolve radically. With $g_0 =10$, for instance, the activity 
range covers typically 12 orders of magnitude from $10^{-6} e^{\bar g}$
to $10^6 e^{\bar g}$ (see \eq{e10}).
If $g_0$ is moderately large, the mutant activities span just a few 
orders of magnitude. This regime is broad enough to permit significant
changes, but not too broad to avoid producing too many lethal changes.
With $g_0 = 3$, for instance, the activity range covers typically $3-4$ 
orders of magnitude from $10^{-1.8} e^{\bar g}$ to $10^{1.8} e^{\bar g}$.
If $g_0$ is small, the lognormal distribution peak can be approximated 
by a Gaussian \cite{RDB2003}
\begin{equation}
\PG(\piA) \simeq \frac{1}{\sqrt{2\pi} g_0 e^{\bar g}} \ 
\exp \left[ \frac{-(\piA - e^{\bar g})^2}{2(g_0 e^{\bar g})^2} \right] .
\end{equation}
The distribution is now narrow and the ranges of values is typically
$[ e^{\bar g}(1-2g_0), e^{\bar g}(1+2g_0) ]$. This type of distribution is 
not adapted for producing large changes, but rather for performing 
fine tuning optimization. With $g_0 = 0.1$, for instance, the activity 
range covers only $\pm 20 \%$ around $e^{\bar g}$.

We remark that the group I ribozyme which we have analyzed corresponds to 
$g_0 \simeq 2.9$, right in the critical regime of moderately large
$g_0$. One can guess from experimental studies of other biopolymers 
or from chemical considerations
that most biopolymers will fall in this range since $\delta G_0$ is 
typically on the order of a few kilocalories while $RT$ is $\simeq 0.6$ kcal
(Note that
$\delta G_0$ corresponds to the free energy change between the biopolymer
native 3D state and an unfolded state, in which the biopolymer has lost
its three-dimensional shape but not its full secondary structure). It
would be interesting to perform further statistical data analysis to see
how, \eg the available protein mutagenesis studies fit with our present
model.

\begin{figure}
\includegraphics[scale=0.33,angle=-90]{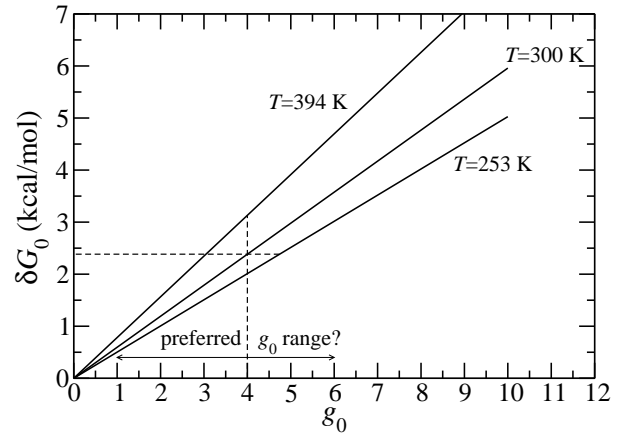}
\caption{Free energy dispersion $\delta G_0$ as a function of $g_0$
for different temperatures $T$. The upper and lower temperature limits for 
life are, respectively, $\simeq 121^\circ$° C and $\simeq~-20^\circ$~C.
Depending on whether the energy statistics is determined by the biophysics or
by evolutionary requirements, the range of either $\delta G_0$ or of $g_0$
is fixed (see for example the dashed lines). Evolutionary requirements 
suggest $1 \lesssim g_0 \lesssim6$.
}
\label{fig6}
\end{figure}

The energy statistics associated mutations is likely to be determined 
at gross scale by the
basic biophysics of the molecules involved. This fixes a range for 
$\delta G_0$. It is nonetheless plausible and suggested by our discussion
that there is an evolutionary preferred type of activity distribution,
and hence of sequences,
that may imply a fine tuning of $g_0 = \delta G_0 /RT$ within the constraints
on $\delta G_0$ coming from biophysics (see \fig{fig6})
so that each mutation typically generates a significant, but not 
systematically lethal, activity change. If one considers that the 
activity changes must cover between, say, one and seven orders of magnitude, 
then the allowed $g_0$ range is $1-6$ (see \eq{e10}). 

To answer the question whether the energy statistics is solely dictated 
by molecular biophysics or whether it is also influenced by evolutionary 
requirements, one may compare the energy statistics of molecules from
different thermal environments. The conservation of the $\delta G_0$ range 
across psychrophilic and thermophilic molecules would stress the domination
of biophysics factors. Note that our model would then imply different 
stochastic evolutionary dynamics, through the width of the activity 
distribution, for psychrophilic and thermophilic
environments. Conversely, the conservation of $g_0$ would reveal 
the importance of evolutionary requirements.

\section{Conclusions}
In this paper, we have presented a model for the distribution of biopolymer
activities resulting from mutations of a given sequence. The 
model is characterized by the statistics of the energy differences
between active conformations and inactive conformations. 
A similar model would be obtained by considering the statistics
of activation energies.
The model fits the measured activity distribution of a ribozyme 
with energy parameters
in the physically appropriate range. It is also able to reproduce 
commonly observed behaviours such as all or none. 

Importantly, the peak of small 
activities exhibits three distinct types depending on the broadness of the
distribution of energy differences. Real biopolymers are in a critical 
regime allowing the exploration of different ranges of activities in a 
few mutations without being too often lethal. This critical regime seems 
the most favorable evolutionary regime and could be the statistical engine
allowing molecular evolution. Thus the present work supports the idea that,
for evolution to take place, 
the temperature and the physico-chemistry dictating the free energy scales
of biopolymers must obey a certain ratio.
At last, it suggests that, by looking at small variations of this ratio, 
one might be able to classify biopolymers. 
One expects, for instance, that biopolymer sequences that are
locked in a shape with a specific function, will have smaller $g_0$
than rapidly evolving biopolymers sequences that could acquire
new functions by undergoing major structural changes. Thus, 
at the origin of life or during rapidly evolving punctuations,
biopolymers with larger $g_0$ than those characterizing highly
optimized, modern RNA and protein molecules,
could have contributed to the emergence of novel phenotypes,
leading thus to an increase of complexity.

\bibliographystyle{apsrev}
\bibliography{bibFB}

\end{document}